\documentclass[showpacs,10pt,twocolumn,prb]{revtex4-1}
\usepackage{amsmath}
\usepackage{amssymb}
\usepackage{graphics}
\usepackage{epsfig}
\usepackage{CJK}
\usepackage{color}

\begin{document}

\begin{CJK*}{GBK}{}

\title{Superconducting state in the metastable binary bismuthide Rh$_{3}$Bi$_{14}$ single crystals}
\author{Xiao Zhang, Hechang Lei, and C. Petrovic}
\affiliation{Condensed Matter Physics and Materials Science Department, Brookhaven
National Laboratory, Upton, NY 11973, USA}
\date{\today}

\begin{abstract}

We report detailed magnetic, transport and thermodynamic properties of metastable Rh$_{3}$Bi$_{14}$ single crystals in superconducting and normal state. We show that Rh$_{3}$Bi$_{14}$ is nearly isotropic, weak to intermediately coupled BCS superconductor, whereas the electronic resistivity above superconducting $T_{c}$ = 2.94 K is dominated by the phonon scattering in the large unit cell with pores filled by Bi atoms. Superconductivity is strongly influenced by the nature of atoms that fill the voids in the crystal structure.

\end{abstract}

\maketitle
\end{CJK*}

\section{Introduction}

Complex crystal structures with atomic subunits often show significant
electronic and magnetic tunability through incorporation of various elements
in the (sub)structure. Examples include covalently bonded IV group
thermoelectric clathrates that are under influence of anharmonic lattice
vibrations with electron - phonon coupling strength comparable to MgB$_{2}$,
alkali doped fullerenes and magnetic molecules.$\cite%
{Crespi,Nolas,Blake,Yokoya,Bryan,Grosche,Sales,Muller}$ Cluster units may
substantially define physical properties whereas their size may considerably
deviate from the length scales defined by the lattice parameters, even in
intermetallic systems where delocalized metallic bonds are present.$\cite%
{Urban}$

Binary Rh-Bi intermetallic phase diagram has been studied for more than
eight decades. Yet, structurally complex metastable Rh$_{3}$Bi$_{14}$ with
136 atoms per unit cell has been discovered only in the recent years.$\cite%
{Rode,Franz,RhBiBr}$ Bismuth-rich side of this phase diagram includes also $%
\alpha $-RhBi$_{2}$, $\beta $-RhBi$_{2}$, RhBi$_{3}$ and RhBi$_{4}$ phases.$%
\cite{Arne,MRuck}$ According to the Rh-Bi phase diagram, Rh$_{3}$Bi$_{14}$
is a metastable phase hidden in the RhBi$_{4}$-Bi solid phase region. Since
the kinetic of peritectic reaction forming RhBi$_{4}$ is extremely slow, it
can be easily suppressed by undercooling due to nucleation problems.$\cite%
{Franz}$ Hence, Rh$_{3}$Bi$_{14}$ single crystals can be decanted from the
Bi liquid whereas the RhBi$_{4}$ single crystals are difficult to obtain.

Rh$_{3}$Bi$_{14}$ has orthorhombic symmetry (Fddd space group) with a =
0.69041(4) [0.68959(15)] nm, b = 1.73816(9) [1.7379(3)] nm, and c =
3.1752(2) [3.1758(6)] nm.$\cite{Franz,RhBiBr}$ In a Rh$_{3}$Bi$_{14}$ unit
cell, there are 24 Rh and 112 Bi atoms.$\cite{RhBiBr}$ Each Rh atom is
surrounded by eight Bi atoms, however there are two inequivalent Rh
positions in the unit cell. Rh1 is cubically coordinated, whereas Rh2
exhibits a square antiprismatic coordination as shown in Fig. 1(a). Each
[RhBi$_{8/2}$] polyhedra shares common edge with four other polyhedra, and
thus create a porous three-dimensional (3D) framework with channels along
the [100] direction. In these channels, \ there are chiral inner surfaces
and filled with bismuth ions (Fig. 1(b)). Partial oxidation of RhBi$_{4}$
using bromine results in Rh$_{3}$Bi$_{12}$Br$_{2}$, isostructural with Rh$%
_{3}$Bi$_{14}$.$\cite{RhBiBr}$ The difference between Rh$_{3}$Bi$_{14}$ and
Rh$_{3}$Bi$_{12}$Br$_{2}$ is in the atoms in the channels. In Rh$_{3}$Bi$%
_{14}$, Bi atoms fill the channels, whereas they are filled by Br atoms in Rh%
$_{3}$Bi$_{12}$Br$_{2}$.\ The theoretical calculation indicates that the 3D
polyhedral frameworks in Rh$_{3}$Bi$_{14}$ and Rh$_{3}$Bi$_{12}$Br$_{2}$ are
formed by covalently bonded rhodium and bismuth atoms, consistent with the
results of high pressure studies.$\cite{RhBiBr}$ On the other hand, there
are no direct bonds between this 3D polycation framework and anions (Bi$%
^{-1} $ or Br$^{-1}$ in the channels), which leads to the higher
compressibility along $a$ axis when compared to those along $b$ and $c$ axes
mainly controlled by the covalently bonded polycation.$\cite{RhBiBr}$

Magnetization measurement indicates that Rh$_{3}$Bi$_{14}$ is a
superconductor with $T_{c}$ = 2.82(5) K.$\cite{Franz}$ Although the crystal
structure and chemical bonding have been studied,$\cite{Franz,RhBiBr,Ruck}$
the studies on physical properties, especially in connection with
superconductivity are still lacking.$\cite{Franz}$ In this work, we report
detailed analysis of Rh$_{3}$Bi$_{14}$ physical properties in the
superconducting and normal states. We show that electronic scattering in the
normal state is dominated by phonons, whereas in its superconducting state Rh%
$_{3}$Bi$_{14}$ is an isotropic, weakly to intermediately coupled BCS
superconductor.

\section{Experimental}

Single crystals of Rh$_{3}$Bi$_{14}$ were grown by the flux-growth method
with Rh:Bi = 5:95 molar ratio. Rh pieces (99.9 $\%$) and Bi shot (99.9 $\%$)
were weighed, combined into alumina crucible, covered with quartz wool and
sealed into the quartz tube with partial pressure of argon. The quartz tube
was heated to 1000 $^{\circ }C$ held constant for 2 h and then cooled at a
rate of -3.5 $^{\circ }C$/h to 300 $^{\circ }C$ where crystals were
decanted. Single crystals with typical size 0.5$\times $0.2$\times $0.2 mm$%
^{3}$ were obtained. Crystal structure and phase purity were examined by
powder and single crystal X-ray diffraction pattern (XRD) with Cu K$_{\alpha
}$ radiation ($\lambda $ = $0.15418$ nm) using a Rigaku Miniflex X-ray
machine. The structure parameters are extracted by fitting the XRD spectra
using the Rietica software.$\cite{Hunter}$ Crystal was oriented using Bruker
SMART APEX II single crystal X-ray diffractometer. The composition of Rh$%
_{3} $Bi$_{14}$ single crystal was determined by examination of multiple
points on the crystals using energy dispersive X-ray spectroscopy (EDX) in
an JEOL JSM-6500 scanning electron microscope. Electrical transport, heat
capacity, and magnetization measurements were performed in a Quantum Design
PPMS-9 and MPMS XL 5.

\section{Results and discussion}

\begin{figure}[tbp]
\centerline{\includegraphics[scale=0.5]{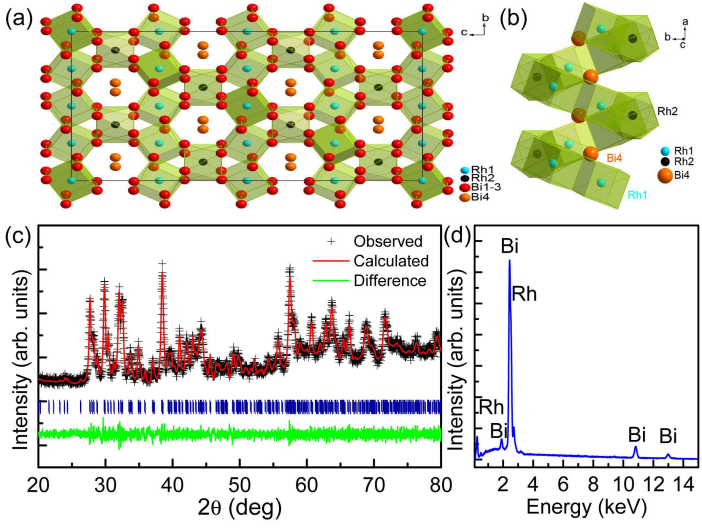}} \vspace*{-0.3cm}
\caption{(\textrm{a}) Crystal structure of Rh$_{3}$Bi$_{14}$.
Three-dimensional framework is built of [RhBi$_{8/2}$] cubes and square
antiprisms which share common edges. Bi4 atoms fill the channels of the
framework. (\textrm{b}) Edge-sharing [RhBi$_{8/2}$] cubes and square
antiprisms enclosing bismuth ions form the zigzag channel in Rh$_{3}$Bi$%
_{14} $. (\textrm{c}) Powder XRD pattern of Rh$_{3}$Bi$_{14}$ single
crystal. (\textrm{d}) The EDX spectrum of a single crystal. }
\end{figure}

All powder X-ray reflections (Fig. 1(c)) can be indexed in the $Fddd$ space
group. The refined lattice parameters are a = 0.6891(1) nm, b = 1.7388(1)
nm, c = 3.1718(2) nm with R$_{p}$ = 5.220, R$_{wp}$ = 6.720 and $\chi ^{2}$
= 0.919, consistent with the values reported in literature.$\cite%
{Franz,RhBiBr}$ On the other hand, we also refined the lattice parameters
using single crystal XRD. The lattice parameters are a = 0.707(3) nm, b =
1.755(5) nm and c = 3.154(8) nm, which are also close to the result obtained
from powder XRD and previous results. EDX spectrum of a single crystal (Fig.
1(d)) confirms the presence of only Rh and Bi with the average atomic ratios
Rh:Bi = 3.0(4):13.8(4).

\begin{figure}[tbp]
\centerline{\includegraphics[scale=0.7]{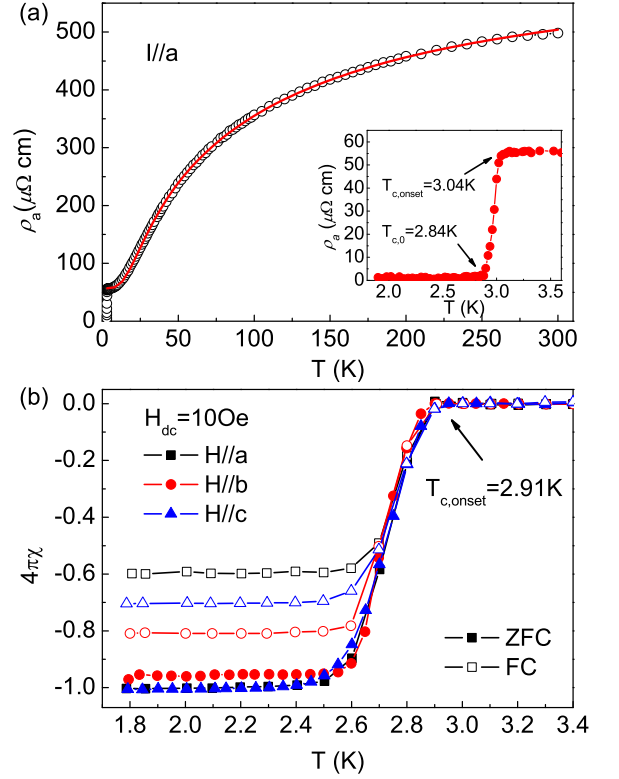}} \vspace*{-0.3cm}
\caption{(\textrm{a}) Temperature dependence of the resistivity $\protect%
\rho _{a}(T)$ of Rh$_{3}$Bi$_{14}$. Inset: resistivity near $T_{c}$. (%
\textrm{b}) Temperature dependence of dc magnetic susceptibility of Rh$_{3}$%
Bi$_{14}$ with ZFC and FC at 10 Oe for $H\parallel $a, $H\parallel $b and $%
H\parallel $c.The superconducting transition temperature $T_{c,onset}$ are
marked by arrow.}
\end{figure}

The temperature dependent electrical resistivity $\rho (T)$ of Rh$_{3}$Bi$%
_{14}$ is shown in Fig. 2(a). The sharp resistivity drop with $T_{c,onset}=$
3.04 K is caused by superconducting transition (inset of Fig. 2(a)). The
residual resistivity ratio(RRR), defined as $\rho (295K)/\rho (3.04K)$, is
about 9.3. The curve $\rho (T)$ is convex above 50 K, with a tendency to
saturate at high temperature. The saturation of $\rho (T)$ could be related
to the Ioffe-Regel limit,$\cite{Ioffe}$ when the charge carrier mean free
path is comparable to the interatomic spacing and/or to the two-band
conductivity.$\cite{Zverev}$ According to the phenomenological model:$\cite%
{GB}$
\begin{equation}
\frac{1}{\rho (T)}=\frac{1}{\rho _{ideal}}+\frac{1}{\rho _{sat}},with\rho
_{sat}\approx \frac{\upsilon _{F}}{\varepsilon _{0}\Omega _{p}^{2}a}
\end{equation}

where $\upsilon_{F}$ is the Fermi velocity, $\Omega_{p}$ is the plasma
frequency, $a$ is interatomic spacing and $\rho_{ideal}=\rho_{r}+\rho_{i}(T)$%
. The $\rho_{i}(T)$ is the inelastic resistivity due to the phonon
contribution whereas $\rho_{sat}$ is the saturation resistivity. Phonon
contribution can be explained by the Bloch-Gr\"{u}neisen formula:
\begin{equation}
\rho_{i}(T)=(\frac{c}{\Theta_{D}})(\frac{T}{\Theta_{D}})^{5}\int^{%
\Theta_{D}/T}_{0}\frac{x^{5}}{(e^{x}-1)(1-e^{-x})}dx
\end{equation}

where $\Theta _{D}$ is the Debye temperature, $c$ is a constant which
depends on the electronic structure of the metal through the Fermi velocity $%
v_{F}$ and the density of states at the Fermi energy.$\cite{JPSJ}$ The
fitting result over the full temperature range of $\rho _{T}$ is shown in
Fig. 2(a) as a solid line, revealing an excellent agreement between the
model and data for $\rho _{r}=$ 60.0(4) $\mu \Omega \cdot $cm, $\rho _{sat}=$
640(2) $\mu \Omega \cdot $cm, $\Theta _{D}=$ 95.0(8) K, $c=$ 27.7(6)$\times $%
10$^{4}$ $\mu \Omega \cdot $cm.

To confirm the presence of bulk superconductivity in Rh$_{3}$Bi$_{14}$
single crystals, the magnetization is measured using dc susceptibility
method. Fig. 2(b) shows the temperature dependence of the dc susceptibility
for a crystal with dimension of 0.2$\times $0.4$\times $0.2 mm$^{3}$ for $%
H\parallel a$, $H\parallel b$ and $H\parallel c$ with zero-field cooling
(ZFC) and field cooling (FC) under an applied magnetic field of 10 Oe. The
steep transition show that the Rh$_{3}$Bi$_{14}$ single crystal has a
superconducting transition temperature for all field directions of 2.91 K($%
T_{c,onset}$) with a transition width $\Delta T_{c}=$ 0.30 K, consistent
with the result of resistivity measurement. At 2 K, the ZFC dc
susceptibility approaches a value of -1 for all directions after
demagnetization correction, indicating bulk superconductivity. The FC signal
shows a flux exclusion of 61 $\%$, 70 $\%$ and 81 $\%$ for $H\parallel a$, $%
H\parallel b$ and $H\parallel c$ respectively. The big superconducting
volume fraction for FC indicates the rather weak flux pinning effects.

\begin{figure}[tbp]
\centerline{\includegraphics[scale=0.7]{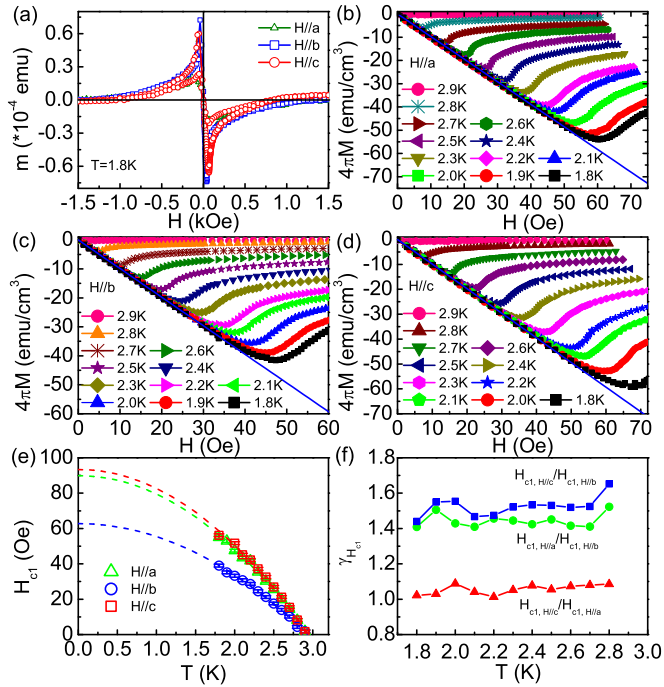}} \vspace*{-0.3cm}
\caption{(\textrm{a}) Magnetization hysteresis loops of Rh$_{3}$Bi$_{14}$ at
T = 1.8 K for $H\parallel $a, $H\parallel $b and $H\parallel $c. (\textrm{b}
), (\textrm{c}) and (\textrm{d}) Low field parts of M(H) at various
temperature for $H\Vert $a, $H\Vert $b and $H\Vert $c with demagnetization
correct, respectively. The solid blue lines are the \textquotedblleft
Meissner line\textquotedblright\ as discussed in the text. (\textrm{e})
Temperature dependence of $H_{c1}$ for $H\Vert $a, $H\Vert $b and $H\Vert $
c. The dashed lines are the fitted lines using $H_{c1}$ = $H_{c1}$(0)(1-($%
T/T _{C}$)$^{2}$).(\textrm{f}) the temperature dependence of anisotropy of $%
H_{c1}$, $\protect\gamma _{_{H_{c1}}}$ = $H_{c1,a}$(T)$/$$H_{c1,b}(T)$, $%
\protect\gamma _{_{H_{c1}}}$ = $H_{c1,c}$(T)$/$$H_{c1,b}(T)$, $\protect%
\gamma _{_{H_{c1}}}$ = $H_{c1,c}$(T)$/$$H_{c1,a}(T)$.}
\end{figure}

The shape of M(H) curve at 1.8 K (Fig. 3(a)) confirms weak pinning, in
agreement with Fig. 2(b). In order to obtain the lower critical fields of Rh$%
_{3}$Bi$_{14}$, we measured the initial $M(H)$ curves at various
temperatures with the field directions along the \emph{a}-, \emph{b}- and
\emph{c}-axis. For each field direction, all curves clearly fall on the same
line and deviate from linearity for different temperature. Linear fits for
the initial parts of magnetization curves describe the Meissner shielding
effects ("Meissner line").

The value of $H_{c1}^{\ast }$ at which the field starts to penetrate into
the sample can be determined by examining the point of deviation from the
Meissner line on the initial slope of the magnetization curve. The first
penetration field $H_{c1}^{\ast }$ is not the same as the real lower
critical field $H_{c1}$, due to the geometric effect. The $H_{c1}$ can be
deduced from the $H_{c1}^{\ast }$, assuming that the magnetization $%
M=-H_{c1} $ when the first vortex enters into the sample. Thus $H$ has been
rescaled to $H=H_{a}-NM$ and $H_{c1}=H_{c1}^{\ast }/(1-N)$ where $N$ is the
demagnetization factor and $H_{a}$ is the external field.$\cite{Fossheim}$
We estimate demagnetization factors 0.47, 0.23 and 0.59 for $H\parallel a$, $%
H\parallel b$ and $H\parallel c$ by using $H_{c1}=H_{c1}^{\ast }$ /tanh($%
\sqrt{0.36b/a}$), where $a$ and $b$ are width and thickness of a plate-like
superconductor.$\cite{Brandt}$ The corrected data are plotted in Fig. 3(b),
(c) and (d). Considering the demagnetization factors, the obtained slopes of
the linear fitting at the lowest temperature of our measurements $T=$ 1.8 K
are -0.972(2), -0.988(2) and -0.987(2), very close to -1 ($4\pi M=-H$) for
all field directions. Thus the full Meissner shielding effect in our
measurement provides a reliable way to determine the value of $H_{c1}$.

$H_{c1}$ is determined as the point deviating from linearity based on the
criterion $\Delta M=(M_{m}-M_{th})=2\times 10^{-6}$ emu, $M_{m}$ is the
measured moment value and $M_{th}$ is fitted moment value at the same
external field. From Fig. 3(b), (c) and (d), $H_{c1}^{a}$, $H_{c1}^{b}$ and $%
H_{c1}^{c}$ values at the different temperatures can be obtained as shown in
Fig. 3(e). The $H_{c1}(T)$ can be fitted according to $%
H_{c1}(T)=H_{c1}(0)[1-(T/T_{c})^{2}]$. Fitting lines are shown in Fig. 3(e).
The obtained $H_{c1}^{a}(0)$, $H_{c1}^{b}(0)$ and $H_{c1}^{c}(0)$ are 90(1),
63(1) and 93(1) Oe respectively. The anisotropy of $H_{c1}$ along each axis
can be determined as shown in Fig. 3(f). It can be seen that the anisotropy
of $H_{c1}(T)$ is very close to 1, especially for $H\parallel c$ and $%
H\parallel a$, indicating that Rh$_{3}$Bi$_{14}$ is a nearly isotropic
superconductor.

\begin{figure}[tbp]
\centerline{\includegraphics[scale=0.8]{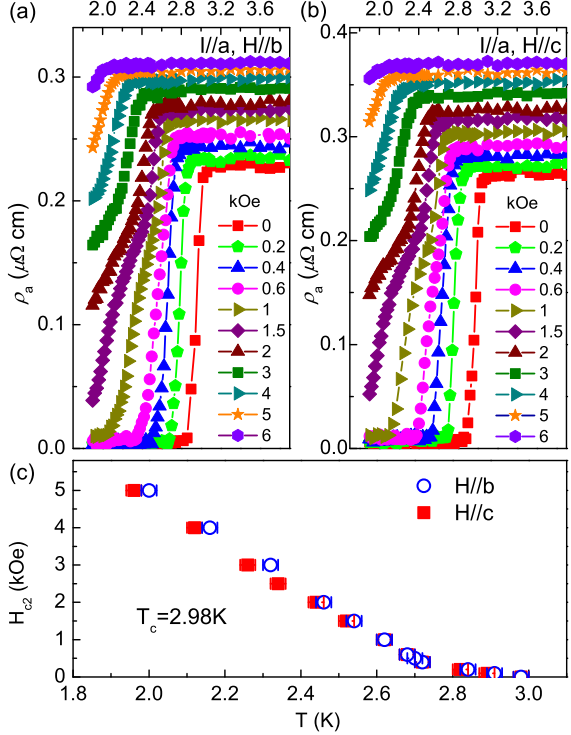}} \vspace*{-0.3cm}
\caption{(\textrm{a}) Temperature dependence of the resistivity in a set of
magnetic fields from 0 to 6 kOe for (a) H$\parallel b$ and (b) H$\parallel c$%
. (c) Temperature dependence of the upper critical field $H_{c2}$ for H$%
\parallel b$ and (b) H$\parallel c$.}
\end{figure}

With increasing magnetic fields (Fig. 4(a) and (b)) the superconducting
transition width broadens and the onsets of transition shift to lower
temperatures gradually for both $H\parallel b$ and $H\parallel c$. When $H=$
6 kOe, the superconducting transition cannot be observed above 1.9 K for
both field directions. It is interesting to observe kinks in the R(T) curves
below $T_{c}$ under fields for both field directions. The kinks are becoming
more pronounced with the increasing field. The origin of these kinks implies
rich vortex physics, similar to NbSe$_{2}$.$\cite{Bhattacharya}$ The upper
critical field $H_{c2}$ is determined by the criterion of 90 $\%$ of the
normal state resistivity at various fields for both field directions (Fig.
4(c)). The slopes $dH_{c2}/dT$ for $H\parallel b$ and $H\parallel c$ are
equal to -6.47(5) kOe/K and -6.01(11) kOe/K, respectively. For dirty limit
superconductors, $H_{c2}(0)$ can be obtained from the
Werthammer-Helfand-Hohenberg formula:$\cite{Werthamer}$
\begin{equation}
H_{c2}(0)=-0.693\frac{dH_{c2}}{dT}\mid _{T_{c}}T_{c}
\end{equation}%
yielding $H_{c2}^{b}(0)$ = 13.4(1) kOe and $H_{c2}^{c}(0)$ = 12.4(2) kOe.
Since the Pauli limiting field $H_{p}(0)=18.4T_{c}$ $\sim $ 52 kOe,$\cite%
{Clogston}$ the orbital effect should be the dominant pair-breaking
mechanism. The small difference in the $H_{c2}$ values for both field
directions indicates that Rh$_{3}$Bi$_{14}$ shows almost isotropic $%
H_{c2}(T) $ as seen in Fig. 4(c).

Since $H_{c1}^{a}\approx $ $H_{c1}^{c}$, according to the anisotropic
Ginzburg-Landau (GL) theory, we assume $\xi _{a}\sim \xi _{c}$ and coherence
length $\xi (0)$ can be estimated from the $H_{c2}(0)$ with: $%
H_{c2}^{b}(0)=\Phi _{0}/[2\pi \xi _{a}(0)\xi _{c}(0)]$ and $%
H_{c2}^{c}(0)=\Phi _{0}/[2\pi \xi _{a}(0)\xi _{b}(0)]$, where $\Phi
_{0}=2.07\times 10^{-15}$ Wb. Based on the values of $H_{c1}(0)$ and $%
H_{c2}(0)$, GL parameters $\kappa _{i}(0)$ is obtained from the equation $%
H_{c2}^{i}(0)/H_{c1}^{i}(0)=2\kappa _{i}^{2}(0)/\ln \kappa _{i}(0)$, where $%
i $ denotes the field applied along $i$ direction. The thermodynamic
critical field $H_{c}(0)$ can be obtained from $H_{c}(0)=H_{c2}^{i}(0)/[%
\sqrt{2}\kappa _{i}(0)]$.The GL penetration lengths are evaluated by the
equations $\kappa _{b}(0)=\lambda _{a}(0)/\xi _{{c}}(0)$ and $\kappa
_{a}(0)\approx \kappa _{c}(0)=\lambda _{a}(0)/\xi _{b}(0)=[\lambda
_{a}(0)\lambda _{b}(0)/\xi _{c}(0)\xi _{b}(0)]^{1/2}$. The anisotropy is $%
\gamma _{anis}=H_{c2}^{b}(0)/H_{c2}^{c}(0)=\xi _{b}(0)/\xi _{{c}}(0)$.$\cite%
{Clem}$ All of the obtained parameters are listed in Table 1.

\begin{figure}[tbp]
\centerline{\includegraphics[scale=0.8]{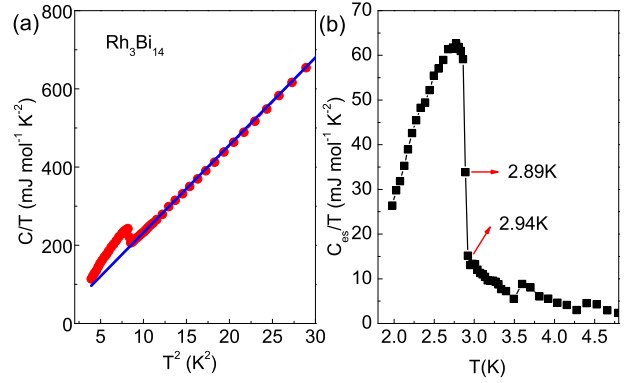}} \vspace*{-0.3cm}
\caption{(\textrm{a}) Low temperature specific heat $C_{p}/T$ $vs.$ $T^{2}$
from 1.95 K to 300 K at $H$ = 0 kOe. The solid line shows the fit (see
text)(b) Temperature dependence of the electronic specific heat of Rh$_{3}$Bi%
$_{14}$ plotted as $C_{es}/T$ $vs$ $T$ at $H$ = 0 kOe. }
\end{figure}

\begin{table*}[tbp]\centering%
\caption{Superconducting parameters of Rh$_{3}$Bi$_{14}$.}
\begin{tabular}{cccccccccccccc}
\hline\hline
Rh$_{3}$Bi$_{14}$ & $T_{c}$ & $H_{c1}^{i}$ & $H_{c2}^{i}(0)$ & $H_{c}^{i}(0)$
& $\kappa _{i}(0)$ & $\xi _{i}(0)$ & $\lambda _{i}(0)$ & $\gamma _{anis}$ & $%
\gamma _{n}$ & $\beta $ & $\lambda _{e-ph}$ & $\Theta _{D}$ & $\Delta
C/\gamma _{n}T_{c}$ \\
& (K) & (Oe) & (kOe) & (kOe) &  & (nm) & (nm) &  & (mJ/mol K$^{2}$) &
(mJ/mol K$^{4}$) &  &  &  \\ \hline
$i=a$ & 2.94 & 90(1) &  &  &  &  & 273(4) & 1.08(3) & 8(1) & 22.39(4) &
0.74(1) & 113.9(7) & 1.8(2) \\
$i=b$ &  & 63(1) & 13.4(1) & 5.4(1) & 17.4(2) & 16.9(2) & 166(8) &  &  &  &
&  &  \\
$i=c$ &  & 93(1) & 12.4(2) & 6.7(1) & 13.1(2) & 15.71(6) &  &  &  &  &  &  &
\\ \hline\hline
\end{tabular}%
\label{TableKey copy(1)}%
\end{table*}%

Specific heat divided by temperature (Fig. 5(a)) shows a jump at 2.94 K,
indicating the bulk superconducting transition. Specific heat includes both
the electron and lattice parts $C(T)=C_{e}(T)+C_{ph}(T)$. In the normal
state, the specific heat of the lattice part is expressed by the $\beta
T^{3} $ term at temperatures far below the Debye temperature $\Theta _{D}$
and electronic specific heat is assumed to be $\sim $ $\gamma T$. Using $%
C_{p}/T=\gamma +\beta T^{2}$ we obtain (Fig. 5(a)) $\gamma =$ 8(1) $mJ$ $%
mol^{-1}$ $K^{-2}$ and $\Theta _{D}$ $\sim $ 113.9(7) K using $\Theta
_{D}=(12\pi ^{4}NR/5\beta )^{1/3}$, where $N=17$ is the number of atoms per
formula unit and R is the gas constant. $\Theta _{D}$ is similar to the
value obtained from $\rho (T)$. The electron-phonon coupling constant $%
\lambda _{e-ph}$ is obtained from the McMillan equation:\newline
\begin{equation}
\lambda _{e-ph}=\frac{\mu ^{\ast }ln(1.45T_{c}/\Theta _{D})-1.04}{%
1.04+ln(1.45T_{c}/\Theta _{D})(1-0.62\mu ^{\ast })}
\end{equation}%
and assuming the common value for the Coulomb pseudopotential $\mu ^{\ast
}\approx $ 0.13. The value of $\lambda _{e-ph}$ is determined to be 0.74(1)
by using $T_{c}=$ 2.94 K and $\Theta _{D}=$ 113.9(7) K. The value of $%
\lambda _{e-ph}$ implies intermediately or weakly coupled BCS
superconductivity. The electronic specific heat $C_{es}$ in the
superconducting state (Fig. 5(b)) is obtained by subtracting the lattice
contribution estimated from the total specific heat. The extracted specific
heat jump at $T_{c}$($\Delta C/\gamma T_{c}=1.8(2)$) is somewhat larger than
the weak coupling value 1.43, and also points to intermediate coupling
strength.$\cite{McMillan}$

Rh$_{3}$Bi$_{12}$Br$_{2}$ has similar structure to Rh$_{3}$Bi$_{14}$, but
the two Bi4 atoms are replaced by the Br atoms. Rh$_{3}$Bi$_{12}$Br$_{2}$ is
a metal above 2 K without superconducting transition.$\cite{Ruck}$ From
theoretical calculation,$\cite{RhBiBr}$ the density of state (DOS) at Fermi
energy is similar for both compounds, however the contributions of atoms (s
and p states of Bi) at Bi4 position to DOS are nearly removed when voids in
the crystal structure are filled with Br atoms. Therefore, the
superconductivity in Rh$_{3}$Bi$_{14}$ might be related to the s and p
states of Bi atom in the interstitial Bi4 position of Rh$_{3}$Bi$_{14}$.

\section{Conclusion}

In summary, we present a comprehensive study of the normal and
superconducting state properties of metastable intermetallic superconductor,
Rh$_{3}$Bi$_{14}$. The temperature dependence of resistivity $\rho(T)$ above
the superconducting transition can be explained by the phonon scattering. Rh$%
_{3}$Bi$_{14}$ is an isotropic, weakly to intermediately coupled BCS
superconductor. Presence (absence) of superconductivity in Rh$_{3}$Bi$_{14}$
(Rh$_{3}$Bi$_{12}$Br$_{2}$) suggests that the nature of atoms that fill the
pores in the structure is rather important.

\section{Acknowledgments}

We thank Kefeng Wang for useful discussions and John Warren for help with
SEM measurements. This work was performed at Brookhaven National Laboratory
and supported by the US DOE under Contract No. DE-AC02-98CH10886.

\end{document}